\documentstyle[aps,prl,epsf]{revtex}

\begin{document}

\twocolumn[\hsize\textwidth\columnwidth\hsize\csname
@twocolumnfalse\endcsname

\title{Complex phase diagram from simple interactions in a one-component system}
\author{E. A. Jagla}
\address{Centro At\'omico Bariloche\\
Comisi\'on Nacional de Energ\'{\i}a At\'omica\\
8400 S. C. de Bariloche, R. N., Argentina}
\maketitle

\begin{abstract}

The pressure-temperature phase diagram of a  one-component system, with  particles
interacting through a spherically symmetric 
pair potential is studied. It is shown that if the pair potential allows for a discontinuous
reduction of the volume of the system when pressure is increased (at zero temperature),
then the phase diagram obtained has
many ``water like'' characteristics. Among these 
characteristics the negative thermal expansion and expansion upon freezing in some 
range of pressure, and the existence of
many different crystaline structures are clearly observable in numerical simulations. 
This implies that complex phase diagrams are not
necessarily originated in complex microscopic interactions.
\end{abstract}

\vskip2pc] \narrowtext

Determination of the phase structure of real materials from first principles
calculations has been one of the aims of statistical mechanics since long
ago. Although a qualitative understanding of the processes leading to the
different kinds of phase transitions (between gas, liquid, and one or more
solid phases) in the pressure-temperature ($P$-$T$) phase diagram of a
classical system has been gained\cite{gen1}, it was clear that
the quantitative fitting of the behavior of real materials would require a
detailed knowledge of the interaction between particles, and a great deal of
computational work, that only in recent years has become feasible. It is
generally assumed that the complexity of the phase diagram of a pure
substance is an indication of the complexity of the elemental interactions
between the atoms or molecules in the system. In this sense the example of
water is paradigmatic\cite{aqua}. Its phase diagram is one of the most
complicated among pure substances, and it includes more than six solid
phases (crystaline or amorphous)\cite{iceamorfo} and a liquid phase with
many properties that are still not well understood from a theoretical point
of view. This complex behavior is believed to be largely due to the
structure of the water molecule, mainly its directional properties.

Although the macroscopic behavior of a material (in particular its
phase diagram) should in principle be derivable from the microscopic
interactions, the assumption that complex macroscopic behavior is due to
complex microscopic interactions may be not true in general. In fact, the
recent developments in the study of complex systems have shown that many
spatially extended models have complex high level behavior which is not an
immediately derivable consequence of its microscopic dynamics, but the
result of frustration between competitive factors at the microscopic scale.
Examples may be taken from the fields of boolean networks, spin glasses, pattern
formation, etc.

In this letter I address the issue of determining the phase behavior of a
pure one-component system, with a particularly chosen spherically symmetric
two-body interaction potential. The first motivation of the work, as already
stated, is to show that rather intricate behavior may appear from simple
microscopic interactions. The second one is the fact that particles
interacting through potentials that may be chosen at will at a great extent
are becoming available in the form of colloidal suspensions of small
spheres, with the aggregate of some amount of non-adsorbing polymer, which
modifies the interaction potential between the spheres\cite{depletion}. So
it is of a basic importance to know the phase behavior of different model
systems, in order to be able to compare the theoretical predictions with the
experimental results.

Whereas for the usual interaction between atoms, consisting of a repulsive
term plus a van der Waals attractive term, a standard solid-liquid-gas phase
diagram develops\cite{generalvdv}, the choice of an interaction $U\left(
r\right) $ with peculiar properties changes the problem in a great extent,
and makes the phase diagram much reacher. The main characteristic of the
interaction from which a complex structure will emerge is the following: $%
U\left( r\right) $ has a range of values in $r$ for which $\partial
U\left( r\right) /\partial r<0$ and $\partial ^{2}U\left( r\right) /\partial
r^{2}\leq 0$. This will imply that when the external pressure $P$ is
increased at $T=0,$ the system will have a discontinuous jump in the mean
interparticle distance at a critical pressure$.$ The detailed form of the
potential beyond this characteristic is of not great importance for the
qualitative features of the phase diagram that is obtained.

The model interaction $U\left( r\right) $ between particles that will be
used consists of a hard core repulsion at a radius $r_{0}$ ($U\left(
r\right) |_{r<r_{0}}=\infty $), is zero for distances larger than a value $%
r_{1}$, and has a soft repulsive part for $r_{0}<r<r_{1}$ of the form $%
U\left( r\right) =\varepsilon _{0}\left( r_{1}-r\right) /\left(
r_{1}-r_{0}\right) $. Two particles interacting through this potential in
the presence of an external force $f$ trying to bring them together, will
have a jump in the interparticle distance from $r_{1}$ to $r_{0}$ when $f$
exceeds the critical value $\varepsilon _{0}/\left( r_{1}-r_{0}\right) .$ If
temperature is measured in units of the energy at contact $\varepsilon _{0}$
(Boltzmann constant is taken to be 1), and distances in units of the hard
core distance $r_{0}$, then $\alpha =r_{1}/r_{0}$ is the only free parameter
of the interaction potential.

\begin{figure}
\narrowtext
\epsfxsize=3.3truein
\vbox{\hskip 0.05truein
\epsffile{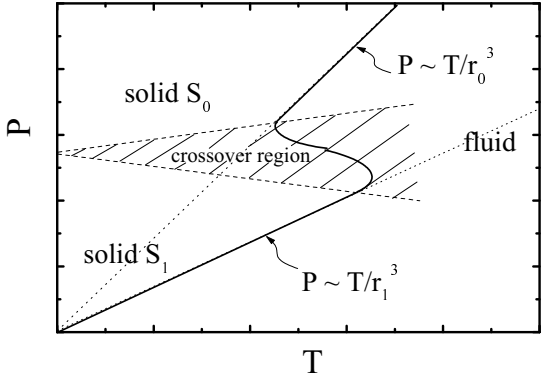}}
\medskip
\caption{Full line: expected behavior for the $P$-$T$ phase diagram of a
model as the one described in text. The two doted lines with $P/T=$constant
are the phase boundaries between solid and fluid phases for hard spheres of
radios $r_{0}$ and $r_{1}$. Within the dashed region a crossover
between a region dominated by the hard core at $r_{0}$ (for higher
temperatures) and a region dominated by an effective hard core at $r_{1}$
(for lower temperatures) should occur.}
\label{f1}
\end{figure}

Well known results on the physics of hard spheres systems\cite{gen1} allow
to gain some insight on the behavior of our model in some limiting regions
of the $P$-$T$ phase diagram: For temperatures much grater than $\varepsilon
_{0}$, the soft repulsive potential is unimportant, and the phase behavior
should correspond to that of hard spheres with radius $r_{0},$ i.e. the
system is in a fluid phase at low pressures and solidificates in a compact
structure (referred to as $S_{0}$) with lattice parameter $r_{0}$ for $%
P\gtrsim P_{0}\equiv T\delta /v_{CP}$, where $v_{CP}$ is the volume per particle at close
packing, and $\delta$ a numerical constant with a value $\delta \simeq
8.3$. For $T\ll \varepsilon _{0}$, particles interact with each other as
hard spheres of radius $r_{1}$ if the pressure is small, i.e., a fluid to solid
transition occurs at $P_{1}\cong P_{0}\alpha ^{-3}$. This solid phase has
lattice parameter $r_{1}$ (and will be referred to as the $S_{1}$
structure). The crossover at intermediate temperatures is expected to occur
when the temperature overcomes the energy difference between the two compact
structures
at the given pressure. This gives the crossover region $P\gtrsim v_{CP}^{-1}\left(
\varepsilon _{0}\pm T\right) /\left( \alpha ^{3}-1\right) $. Then the overall
fluid-solid coexistence curve of the system should have an `S' shape as
depicted in Fig. 1. The existence of the zone with a negative value of $%
\partial P/\partial T$ in the coexistence curve will depend essentially on $%
\alpha $, and will occur only for high enough values of this parameter. 
In addition, in the solid phase and within the crossover region the transition
between structures $S_0$ and $S_1$ should occur.

It must be noticed that our model does not contain an attractive part in the
potential, and for this reason it lacks a liquid phase. However a liquid phase 
can be obtained within a generalized van der
Waals approach\cite{generalvdv}, in which an attractive interaction is included
through an energy term of the form $-\gamma /v$, with $v$ being the volume per
particle. This modification does not affect the structure of the phases, and
the nature of the anomalies of the phase diagram that are discussed below.

Numerical simulations of the model in two- and three-dimensions reveal that
the features already described are correct, and also that new
qualitative behavior arises. I
will describe now the result for the phase diagram obtained in the case of a
two dimensional system with $\alpha =1.65.$ This value was chosen in order
that the second neighbors interaction in the most compact structure $S_{0}$
is zero. The results shown correspond to two dimensions in order to make
easier the presentation of the structures that appear, but the qualitative
features are the same in three dimensions with of course different structures%
\cite{nota}. The system consists of 64 particles (it was checked that larger
systems do not introduce new qualitative behaviors), lying in a cage with
periodic boundary conditions. The cage was not square, but slightly
distorted in order to accommodate a compact structure of eight rows of eight
particles each.

Montecarlo simulations were performed at constant pressure in the following
way: after updating the coordinates of all particles to allow for thermal
wandering according to a standard Metropolis algorithm, a trial global
expansion or contraction of the system (with a corresponding rescaling of
all coordinates of the particles) was performed, and accepted with the
Metropolis rule according to the value of the energy change $\Delta E$,
given by $\Delta E=P\Delta V-\left( NT/V\right) \Delta V+dE$, where $N$ is
the number of particles, $V$ the volume of the system, 
$dE$ is the energy change associated to the change
of interparticle distance, and the term $-\left( NT/V\right) \Delta V$ gives
the correct expression in the case of an ideal gas (i.e., when $dE=0$), and
its appearance is a consequence of the use of periodic boundary conditions
(the absence of walls). Different runs were performed at constant pressure
starting from random configurations at high temperature, cooling down to
zero temperature and then warming up. Around 2500 Montecarlo steps were used
for thermalization at each temperature, and then 10000 steps were used to
calculate thermodynamical quantities, such as the equilibrium mean volume $V$
of the system, diffusion coefficient of particles, and internal energy.

\begin{figure}
\narrowtext
\epsfxsize=3.3truein
\vbox{\hskip 0.05truein
\epsffile{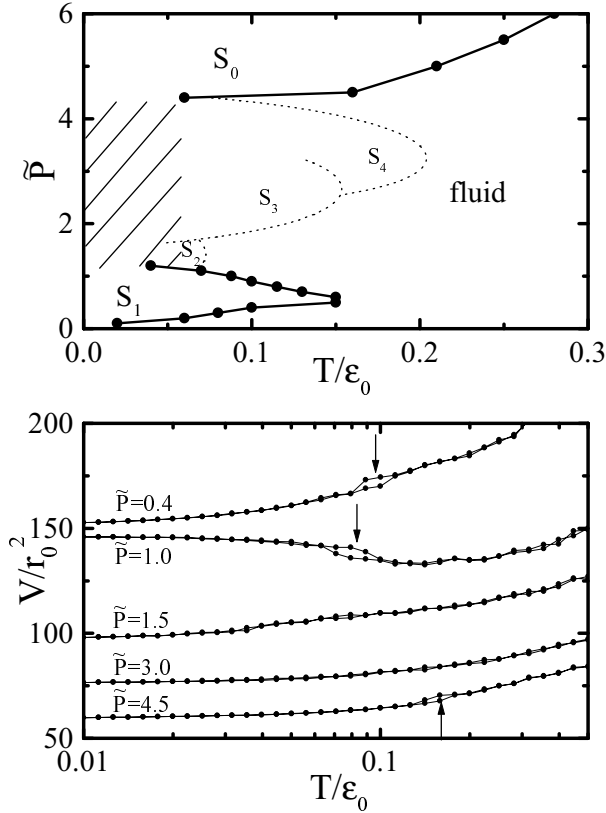}}
\medskip
\caption{(a)Phase diagram obtained from the numerical simulations, for $%
\alpha =1.65$. Continuous lines indicate a first order transition between
solid and fluid. Dashed region is a zone where the system freezes upon
cooling but with no clear evidence of a phase transition. Dotted lines 
are approximate limits of stability of possible 
crystaline fundamental states (see text
for further explanation. Labelling of the structures correspond to that of
Fig. 4). (b) Mean volume of the system for different pressures as a function
of temperature, upon cooling and heating. First order transitions are
indicated by arrows.}
\label{f2}
\end{figure}

The phase diagram obtained for $\alpha =1.65$ is depicted in Fig. 2(a).
Continuous lines for the reduced pressure $\widetilde{P}\equiv
Pr_{0}^{2}/\varepsilon _{0}$ in the ranges $\widetilde{P}$ $\lesssim 1.2$ and 
$\widetilde{P}\gtrsim 4.3$ indicate first order transitions between fluid and
solid. The solid phase has the $S_{0}$ structure for $\widetilde{P}>4.3,$ or 
$S_{1}$ for $\widetilde{P}<1.2$. The isobaric $V$-$T$ curves are shown in
Fig. 2(b). Note the hysteresis loops for small and large pressures, which is
a signature of the first order transition, and also the anomalous
compressibility of the fluid near freezing when $.5<\widetilde{P}<1.2$. In this
region the usual (entropic) volume reduction when temperature is reduced is
overcome by the expansion produced when particles diminish their kinetic
energy and move out of the soft core of their neighbors. In the intermediate
pressure range $1.2<\widetilde{P}<4.3$ no clear first order solidification
transition is observed in the simulations, but a rather continuous 
freezing into disordered
states. One signature of this fact is the diffusion coefficient of the
particles which was found to decrease from a large value in the fluid, to a
very small value near $T\simeq 0.05$ in the intermediate pressure
regime, though no anomalies in the volume or energy were observed. This is
indicated by the shaded region in Fig. 2(a) (the dotted lines and the labels 
$S_{2}$, $S_{3}$ and $S_{4}$ in this figure are discussed below).

\begin{figure}
\narrowtext
\epsfxsize=3.3truein
\vbox{\hskip 0.05truein
\epsffile{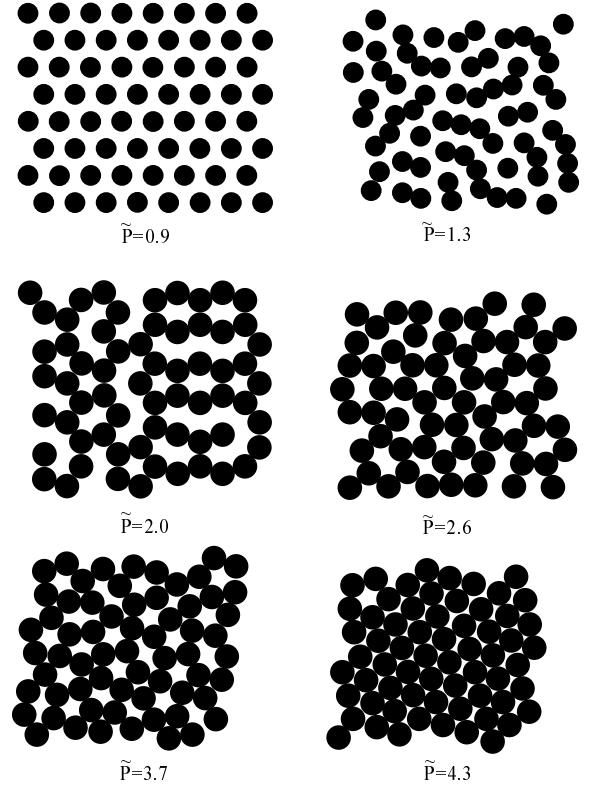}}
\medskip
\caption{Zero temperature configurations obtained when cooling down the
system at different values of the reduced pressure $\widetilde{P}$ 
($\equiv Pr_{0}^{2}/\varepsilon _{0}$). The
radius of the black dot indicates the hard core at $r_{0}$. The value of $%
\alpha $ is 1.65. Some crystaline defects are observable for 
$\widetilde{P}=4.3.$ Note that
between $\widetilde{P}\sim 1.2$ and $\widetilde{P}\sim 4$ the system does
not crystallize.}
\label{f3}
\end{figure}

In Figure 3 snapshots of the states obtained after cooling down to zero
temperature at different pressures are shown. Those corresponding to the
intermediates ranges of pressure show a notable diversity, being striking in
particular the profusion of pentagons for $\widetilde{P}=2.6$ and $%
\widetilde{P}=3.7$. The origin of these structures may be traced back to the
competition between the two terms in the energy of the system at $T=0$: One
is the usual $PV$ term, which tends to minimize the volume, and the other is
the repulsive energy term, which tends to maximize the interparticle
distance. This produces a sort of frustration, because both terms cannot be
minimized at the same time. The two compact structures $S_{0}$ and $S_{1}$
correspond to two ways of reducing the energy by minimizing one term whereas
maximizing the other. These are the best compromises in the case of very low
or very high pressures. However, when both energy terms are comparable,
lower energy intermediate solutions can be found by arranging the particles
with a coordination number (number of neighbors at distance $r_{0}$)
intermediate between 0 and 6 (which correspond to the structures $S_{1}$ and 
$S_{0}$). 

\begin{figure}
\narrowtext
\epsfxsize=3.3truein
\vbox{\hskip 0.05truein
\epsffile{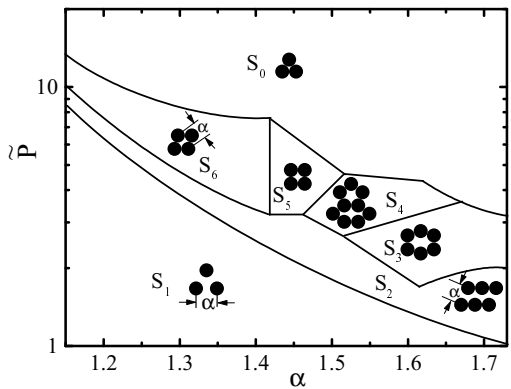}}
\medskip
\caption{Possible fundamental (crystaline) states of the system as a
function of $\widetilde{P}$ and $\alpha .$ Inside each region there is a
sketch of the pattern used to cover up the plane. The structures shown have
lower or equal energy than those found in the numerical simulations, but
additional structures cannot be ruled out.}
\label{f4}
\end{figure}

In fact, different crystaline configurations can be proposed, and their
energy calculated in order to find the most stable one as a function of $%
\widetilde{P}$ and $\alpha $. The result of this analysis is shown in Fig.
4. It is clear that the triangular structures $S_{0}$ and $S_{1}$ are not
the lowest energy states for an intermediate range of pressures. The
proposed structures in Fig. 4 are not necessarily the fundamental states,
because other structures may have been missed. However, the energy of the
proposed structures is lower than the values obtained in the numerical
simulations for all the points simulated in the $\widetilde{P}$-$\alpha $
plane. This figure suggests that in the region $1.2<\widetilde{P}<4.3$ at
low temperatures and for $\alpha =1.65$, at least three different
crystaline structures exist. Starting from these three structures as
initial configurations at zero temperature, numerical runs were performed at
different pressures by increasing and then decreasing the temperature. The
periodic structures persist upon heating up to some maximum temperature. This
temperature and a comparison between the energies of the periodic 
structures and the disordered ones (obtained on cooling) allow to determine 
the stability limit of each structure. The corresponding curves for the three
structures are plotted in Fig. 2(a) as dotted lines.

As it was already said, there may be other thermodynamically stable states
in some regions of the $P$-$T$ plane that were missed. In particular, it
must be said that the profusion of pentagons in the structures of Fig. 3
raises the question of whether in some region of the $P$-$T$ plane, the
lowest energy state of the system may be amorphous or even quasicrystaline.
More work is needed to answer this question but in any case it is worth
noting that the failure from the computational point of view of reaching a
crystaline structure when temperature is decreased in the intermediate
pressure range is (at least) an indication of the existence of many
metastable low energy disordered states.

The same qualitative features of the phase diagram were obtained also in
three-dimensional simulations. The minimum value of $\alpha $ necessary to
get the volume anomaly at melting is about $1.2$, both for two- and
three-dimensional systems. This anomaly in the fluid-solid coexistence, and
the negative thermal expansion coefficient in this region (see Fig. 2(b))
reminds strongly the same effects occurring in water. Although this model
does not intend to be a good description of real water, an interaction
potential with a double minimum (which is related to the potential
considered here) has been suggested to be the origin of the density anomaly
in water\cite{waterprl}. A comparison of the phase diagrams of water or
Bismuth\cite{bismuto} with the one presented here reveals in fact many
coincidences, as the mentioned anomalies and the position of the different
crystaline phases in the $P$-$T$ diagram, which occur near the pressure
where the melting temperature is minimum. In addition, amorphous structures
in ice are well known\cite{iceamorfo,otroice}, and may be related to the
states obtained in the numerical simulation when cooling down at
intermediate pressures. The point to be stressed here is that a complex
phase diagram of a pure substance, as those of water or Bismuth, may
ultimately be originated by the frustration that particular, simple interactions may
introduce in the system.

The author thanks J. Simon\'{\i}n and K. Hallberg for critical reading of the
manuscript. This work was financially supported by Consejo Nacional de
Investigaciones Cient\'{\i }ficas y T\'{e}c\-ni\-cas (CONICET), Argentina.


\begin{references}
\bibitem{nota}  It may be objected that the nature of the melting of a
two-dimensional system is a controversial point. However, for small systems
as the one simulated here the signatures obtained in numerical simulations
correspond to a first order melting transition. As the main reason for using
a two-dimensional system is simply to make clear the discussion of the
results (which apply to three-dimensional experimental situations), the point
is not relevant in this context.
\end{references}
\end{document}